\documentclass[twoside]{ilcws07}
\usepackage[latin1]{inputenc}
\usepackage[dvips]{graphicx,epsfig,color}
\usepackage{wrapfig}
\usepackage{amssymb,amsmath,array}

\begin{document}
\title{The charged Higgs boson mass in the 2HDM: decoupling and CP violation
} 
\author{Maria Krawczyk$^{1,2}$ \thanks{Supported in part by
 EU Marie Curie Research Training Network
HEPTOOLS, under contract MRTN-CT-2006-035505 and by FLAVIAnet
contract No. MRTN-CT-2006-035482.
 } \, and Dorota Soko\l owska$^1$ \\
%
\vspace{.3cm}\\
1- Institute of Theoretical Physics, University of Warsaw \\
00-681 Warsaw, ul. Ho\.za 69, Poland
\vspace{.1cm}\\
2-  TH-Division, CERN, CH-1211 Gen\'eve 23, Switzerland\\
}
\maketitle

\begin{abstract}
Mass range of the charged Higgs boson in the 2HDM with explicit and
spontaneous CP violation is discussed.\,Constraints on $M_{H^\pm}$
in the CP conserving 2HDM(II) are shown.
\end{abstract}

\section{The 2HDM potential and spontaneous symmetries breaking}
The most general, invariant under gauge group $SU(2)_{L} \times
U(1)_{Y}$ and renormalizable potential of the Two Higgs Doublet
Model (2HDM) \cite{Branco:1999fs,haber,Ginzburg:2004vp} is given by
\begin{equation}
\begin{split}
V = \frac{\lambda_{1}}{2} \left(\Phi_{1}^{\dagger}
\Phi_{1}\right)^{2} + \frac{\lambda_{2}}{2} \left(\Phi_{2}^{\dagger}
\Phi_{2}\right)^{2} +\lambda_{3} \left(\Phi_{1}^{\dagger}
\Phi_{1}\right) \left(\Phi_{2}^{\dagger} \Phi_{2}\right) +
\lambda_{4} \left(\Phi_{1}^{\dagger} \Phi_{2}\right)
\left(\Phi_{2}^{\dagger} \Phi_{1}\right) \\+ \left[ \frac{1}{2}
\lambda_{5} \left(\Phi_{1}^{\dagger} \Phi_{2}\right)
\left(\Phi_{1}^{\dagger} \Phi_{2}\right) + \lambda_{6}
\left(\Phi_{1}^{\dagger} \Phi_{1}\right) \left(\Phi_{1}^{\dagger}
\Phi_{2}\right) +  \lambda_{7} \left(\Phi_{2}^{\dagger}
\Phi_{2}\right) \left(\Phi_{1}^{\dagger} \Phi_{2}\right) + h.c.
\right] \\ -\frac{1}{2}m_{11}^{2} \left(\Phi_{1}^{\dagger}
\Phi_{1}\right) -\frac{1}{2}m_{22}^{2} \left(\Phi_{2}^{\dagger}
\Phi_{2}\right) -\left[ \frac{1}{2}m_{12}^{2}
\left(\Phi_{1}^{\dagger} \Phi_{2}\right) + h.c. \right],
\label{potencjal_ogolny}
\end{split}
\end{equation}
where $\lambda_{1-4},m_{11}^{2},m_{22}^{2} \in \mathbb{R}$ (by the
hermicity of the potential), while in general  $\lambda_{5-7},
m_{12}^{2} \in \mathbb{C}$. In the most general CP breaking form it
has 14 parameters, however  only 11 are independent, see e.g.
\cite{Lavoura:1994yu,Accomando:2006ga}.
 In the model there are five
 Higgs particles: three neutral  $h_1,h_2,h_3$ (for CP
conservation - two CP-even $h,H$ and one CP-odd $A$) and two charged
Higgs bosons $H^\pm$.
\subsection{$Z_2$ and CP symmetries}
The $Z_{2}$ symmetry of the potential (\ref{potencjal_ogolny}) is
defined as the invariance of $V$ under the following transformation
of doublets: $ \Phi_{1} \to -\Phi_{1}, \Phi_{2} \to \Phi_{2} \quad
\textrm{or} \quad \Phi_{1} \to \Phi_{1}, \Phi_{2} \to - \Phi_{2}$.
 If $Z_2$ (in either form)  is a symmetry of the potential, then $m_{12}^{2}
=~\lambda_6 =~\lambda_7 =~0$. The $Z_2$ symmetry is {\it{softly}}
broken by the terms proportional to $m_{12}^{2}$.

General 2HDM allows for CP violation both explicitly and
spontaneously  \cite{Lee:1973iz,Branco:1979pv,Branco:1999fs}. The
CP violation can be naturally suppressed by imposing   a $Z_2$
symmetry on  the Higgs potential.

\subsection{Reparametrization transformation}

A global unitary transformation which mix two doublets and change
their relative phase does not change the physical content of 2HDM as
discussed recently in \cite{ivanov}, see also
\cite{haber,Ginzburg:2004vp,Branco:1999fs}.
 It is given by
\begin{eqnarray}
\label{trans_repar_macierz} \left(\begin{array}{c}  \Phi'_{1}\\ \Phi'_{2}\\  \end{array}\right) = \mathcal{F} \left(\begin{array}{cc}  \Phi_{1}\\ \Phi_{2}\\  \end{array}\right), \qquad  \mathcal{F} = e^{-i \rho_{0}} \left( \begin{array}{cc} \cos \theta e^{i \rho/2} & \sin \theta e^{i(\tau - \rho/2)} \\
- \sin \theta e^{-i(\tau - \rho/2)} & \cos \theta e^{- i\rho/2} \\  \end{array} \right).
\end{eqnarray}
There are three {\it{reparametrization}} parameters - $\rho, \theta,
\tau$, and in  addition   $\rho_0$ parameter  as an overall phase.
If $\theta = 0$ there is no mixing of two dublets and  the
transformation becomes a global transformation of doublets  with an
independent phase rotations ({\it{rephasing}}):
\begin{eqnarray}
k = 1,2 : \Phi_{k} \to e^{-i \rho_{i}} \Phi_{k}, \quad \rho_{1} = \rho_{0} - \frac{\rho}{2}, \quad \rho_{2} = \rho_{0} + \frac{\rho}{2}, \quad \rho = \rho_{2} - \rho_{1}. \label{rpht_pola}
\end{eqnarray}
The original form of the potential is recovered by the appropriate
changes of phases of the following coefficients:

\subsection{Explicit and spontaneous CP violation in 2HDM}
CP violation may occur in 2HDM only if $Z_2$ symmetry is broken
\cite{Branco:1979pv,Branco:1999fs,haber,Ginzburg:2004vp,ivanov}. A
necessary condition for an {\it{explicit CP violation}} in the Higgs
potential $V$ is an existence of complex parameters. However, if
there exists a reparametrization leading to $V$ with only real
parameters ({\it{real basis}}), then there is no explicit CP
violation in $V$. A spontaneous CP breaking, by the vacuum state, is
still possible \cite{Lee:1973iz,Branco:1979pv,Branco:1999fs}.

In the simply analysis \cite{ks}, which results we present here,
only the potential with exact and softly broken $Z_2$ symmetry  was
considered, i.e. $\lambda_{6,7}=0$. In studying 2HDM with an
{\it{explicit}} CP conservation or violation the {\it {real vacuum
representation}} \cite{Ginzburg:2004vp} was applied. A spontaneous
CP violation was discussed assuming the {{\it explicitly CP
conserving V}}.

\subsection{Vacuum expectation values}

The most general vacuum (extremum) state can be described by
\cite{Branco:1979pv,DiazCruz:1992uw,Barroso:2007rr,Ginzburg:2007jn,ks}
\begin{eqnarray}
\left\langle \Phi_{1} \right\rangle = \frac{1}{\sqrt{2}}
\left( \begin{array}{c} 0\\ v_{1} \\ \end{array} \right), \qquad \left\langle \Phi_{2}
\right\rangle = \frac{1}{\sqrt{2}} \left( \begin{array}{c} u\\ v_{2}e^{i \xi} \\ \end{array}
\right), \label{VEVS}
\end{eqnarray}
where $v_{1}, v_{2}, \xi, u \in \mathbb{R}$.  By gauge
transformation one can  always make $v_{1} > 0$.  Below we will
assume that $v_2 \neq 0$, with $v^2=v_1^2+v^2_2=(246
\,\,{\rm{GeV}})^2$, and  $0 \leq \xi < 2 \pi$.

For vacuum with  $u \not=0$  the electric charge is not conserved
and the photon becomes a massive particle ("charged vacuum"). If $u
= 0$ then  a "neutral vacuum" are possible. Depending on the value
of $\xi$ there may or may not be a spontaneous CP violation
\cite{Branco:1979pv,haber,Barroso:2007rr,Ginzburg:2007jn}.
 The useful quantity is
$\nu=\frac{m_{12}^2}{2v_1v_2}$ (or $\nu=\frac{\Re
m_{12}^2}{2v_1v_2}$) \cite{Ginzburg:2004vp}, which here is taken to
be positive.

\subsection{Extremum conditions}
For the  extremum states (\ref{VEVS}) the first derivatives of the
considered potential lead to the following set of extremum
conditions:
\begin{eqnarray}
&\label{slabo1} 0 = u \left[ v_{1} v_{2} \cos \xi \left( \lambda_{4}
+ \lambda_{5} \right) - m_{12}^{2} \right], \,\, \label{slabo2} 0 =
u \left[  \lambda_{2} \left(u^{2} + v_{2}^{2}
\right) + \lambda_{3} v_{1}^{2} - m_{22}^{2} \right]\\
& 0 = v_{2} \sin \xi \left[  2 \lambda_{5} v_{1} v_{2}
 \cos \xi  - m_{12}^{2}\right],\,\,  0 = v_{2} \sin \xi \left[ v_{1}^{2}
 \left( \lambda_{3} + \lambda_{4} - \lambda_{5} \right) + \lambda_{2}
  \left(u^{2} + v_{2}^{2} \right) - m_{22}^{2} \right] \\
& \label{slabo6} 0 = v_{1} \left[ v_{2}^{2} \left( \lambda_{5}
\cos^{2}2\xi + \lambda_{4} \right)  + \lambda_{1}
 v_{1}^{2} + \lambda_{3} \left( u^2 + v_{2}^{2}   \right) - m_{11}^{2} \right]- m_{12}^{2} v_{2} \cos \xi \\
& \label{slabo5} 0 = u v_{1} v_{2} \sin \xi \left( \lambda_{4} -
\lambda_{5} \right), \,\, 0 = v_{2} \cos \xi \left[ v_{1}^{2}
\left(\lambda_{3} + \lambda_{4} + \lambda_{5} \right)  + \lambda_{2}
\left(u^{2} + v_{2}^{2} \right) - m_{22}^{2} \right] - m_{12}^{2}
v_{1}.
\end{eqnarray}
If $u = 0$ then above conditions are satisfied  for an exact $Z_2$
symmetry ($m^2_{12}=0$) when  the only possible neutral vacuum state
is the one which respects CP, i.e. with $\sin \xi = 0$,
 and for a broken  $Z_2$ symmetry. In the latter case two neutral vacuum states are possible -
 without and with  CP violation, for   $\sin \xi = 0$ and $\sin \xi \not = 0$,
 respectively. To get a real minimum of the potential  the eigenvalues
of the squared mass matrix have to be positive. We will assume in
addition that positivity constraints hold guaranteeing stability of
the vacuum \cite{Deshpande:1977rw}.

\subsection{Physical regions for CP conserving 2HDM}
Expressions for masses of $H^\pm$ and $A$ for 2HDM with an explicit
or a spontaneous CP {\it{conservation}} are as follows.
 \vspace*{-0.3cm}
\paragraph{$Z_2$ symmetry broken} If $Z_2$ symmetry  is
softly broken ($\nu \neq 0$), then the masses squared of $H^{\pm}$
and $A$ are given by:
\begin{eqnarray}
M_{H^{\pm}}^{2} =  v^{2} \left( \nu - \frac{1}{2} \left(\lambda_4 +
\lambda_5 \right) \right), \quad M_{A}^{2} =  v^{2} \left(\nu -
\lambda_{5} \right). \label{cpha}
\end{eqnarray}
In order to have positive $M^2_{H^{\pm}}$ and $M^2_A$ inequalities
$\lambda_{5} + \lambda_{4} < 2 \nu {\rm{\,\,and\,\,}} \lambda_{5} <
\nu$ should hold.

Large masses for $H^\pm$ and $A$ (\ref{cpha}) can arise from large
$\nu$. In the limit $\nu \to \infty$ the decoupling is realized -
$h$ is like the Higgs boson in the Standard Model, while $H^\pm, A,
H$ are heavy and almost degenerate \cite{haber,Ginzburg:2004vp}.

\vspace*{-0.3cm}
\paragraph{Exact $Z_2$ symmetry}
The results for an exact $Z_2$ symmetry  can be obtained from above
expressions in the limit $\nu \to 0$. Then $\lambda_5 <0$. Masses
cannot be too large, as here they can arise only due to $\lambda's$.
However, large $\lambda's$ may violate  tree-level unitarity
constraints \cite{Ginzburg:2005dt}.

\subsection{Physical regions for CP violating 2HDM}
As it was mentioned above  if  the 2HDM potential breaks $Z_2$
symmetry then CP violation may be realized in the model.  Note, that
if CP is violated  physical neutral Higgs states are $h_1,h_2,h_3$,
without definite CP properties, while  $h,H,A$ are useful but only
auxiliary states.

\vspace*{-0.3cm}
\paragraph {Explicit CP violation} If there is explicit CP violation
all formulae derived for the CP conservation  case (\ref{cpha} and
beyond)  hold after the replacements: $\lambda_5 \to \Re \lambda_5$
and $m_{12}^2 \to \Re m^2_{12}$. Note, that   the decoupling can be
realized here as well, with  large $M_{H^\pm}^2$  arising from
large $\nu$.

\vspace*{-0.3cm}
\paragraph {Spontaneous CP violation} Spontaneous CP violation
may  appear if  there is  a CP breaking phase of the VEV, so
$\sin\xi \not = 0$. From the extremum condition one gets that:
\begin{eqnarray}
&\cos \xi =
\frac{\displaystyle{m_{12}^{2}}}{\displaystyle{\lambda_{5} 2 v_{1}
v_{2}}} = \frac{\displaystyle{\nu}}{\displaystyle{\lambda_{5}}},&
\label{cosinus_NCP}
\end{eqnarray}
from which it follows that  $|\nu / \lambda_5| < 1$. The  squared
masses for $H^{\pm}$ and $A$ are given by the following expressions,
see also \cite{Ginzburg:2007jn}:
\begin{eqnarray}\label{ha}
M_{H^{\pm}}^{2} =  \frac{v^{2}}{2} \left(\lambda_5 - \lambda_4
\right),
 \quad M_{A}^{2} = \frac{v^{2}}{\lambda_5} \left(
  \lambda_{5}^{2} - \nu^{2} \right)
  = v^{2} \lambda_5 \sin^{2}
 \xi.
\end{eqnarray}
 We see that they are quite different from
the formulae for $M_{H^{\pm}}^{2}$ and $M_A^2$ discussed above.
(Note, that although $A$ is no longer a physical state, positivity
of $M_A^2$  still provides a good constraint since it {gives at the
same time a condition for positivity of squared masses of physical
particles}.)
 From the last expression for $M_{A}^{2}$ (\ref{ha}) it is easy to see that
$\lambda_5$ have to be positive. Furthermore, squared masses
(\ref{ha}) are positive if $\lambda_5> \lambda_4$ and $\lambda_5 >
\nu > 0$.

It is worth mentioning that  the squared mass of $H^{\pm}$ does not
depend on $\nu$ at all. Therefore, $M_{H^{\pm}}$  cannot be too
large in 2HDM with CP  violated spontaneously, for the same reason
as in the discussed above case of  exact $Z_2$ symmetry.

\subsection{Conclusion on possible vacuum states in 2HDM}
Regions where various  vacuum states (conserving or spontaneously
violating CP) can be realized in 2HDM are mutually exclusive
\cite{Deshpande:1977rw,Barroso:2007rr,Ginzburg:2007jn,ks}. The mass
of charged Higgs boson may serve as a guide over various regimes of
the 2HDM. Existence of heavy charged Higgs boson, with mass above
600-700 GeV \cite{Ginzburg:2004vp,ks}, would be a signal that in
2HDM $Z_2$ symmetry is violated, and  CP can be violated only
explicitly.

\section{Experimental constraints on  the 2HDM(II) with CP conservation}
Here we consider  the CP conserving 2HDM, assuming that $Z_2$
symmetry  is extended also on the Yukawa interaction, which allows
to suppress the FCNC \cite{Glashow:1976nt}. We limit ourself to
constraints on the Model II of the Yukawa interaction, as in MSSM,
see e.g. \cite{Krawczyk:2002df}. There are 7 parameters for the
potential with softly breaking $Z_2$ symmetry: masses $M_h,
M_H,M_A,M_{H^{\pm}}$, mixing angles $\alpha$ and $\tan
\beta=v_2/v_1$, and parameter $\nu$.

{Couplings (relative to the corresponding couplings of the SM
Higgs)} are as
follows:\\
\hspace*{8cm} {{\fbox{h}}} \hspace{3cm} {\fbox{A}}\\
{ to W/Z: \hspace{6.cm}{$\chi_V= \sin(\beta-\alpha)$}
\hspace{2cm} {0}\\
to down quarks/charged leptons:   \hspace{0.65cm} {$\chi_d=\chi_V
-\sqrt{1-\chi_V^2} \tan \beta $}
\hspace{1cm} {$-i \gamma_5 \tan \beta $}\\
 to up quarks:\hspace{3.8cm} {$\chi_u=\chi_V +  \sqrt{1-\chi_V^2}/\tan \beta $} \hspace{0.8cm}
 {$-i\gamma_5/\tan\beta $}
}
\vskip 0.1cm

 {$H$} couples like  {$h$} with following
replacements:
 {{$\sin(\beta-\alpha) \to
 \cos(\beta-\alpha)$}} and $\tan \beta \rightarrow -\tan \beta$.
 For large $\tan \beta$ there are  enhanced couplings to $d-$type fermions.
Note, that coupling
 $\chi_{VH^+}^h = \cos(\beta-\alpha)$ is complementary to the $\chi^h_V$.

\begin{figure}[h]
  \includegraphics[height=.3\textheight]{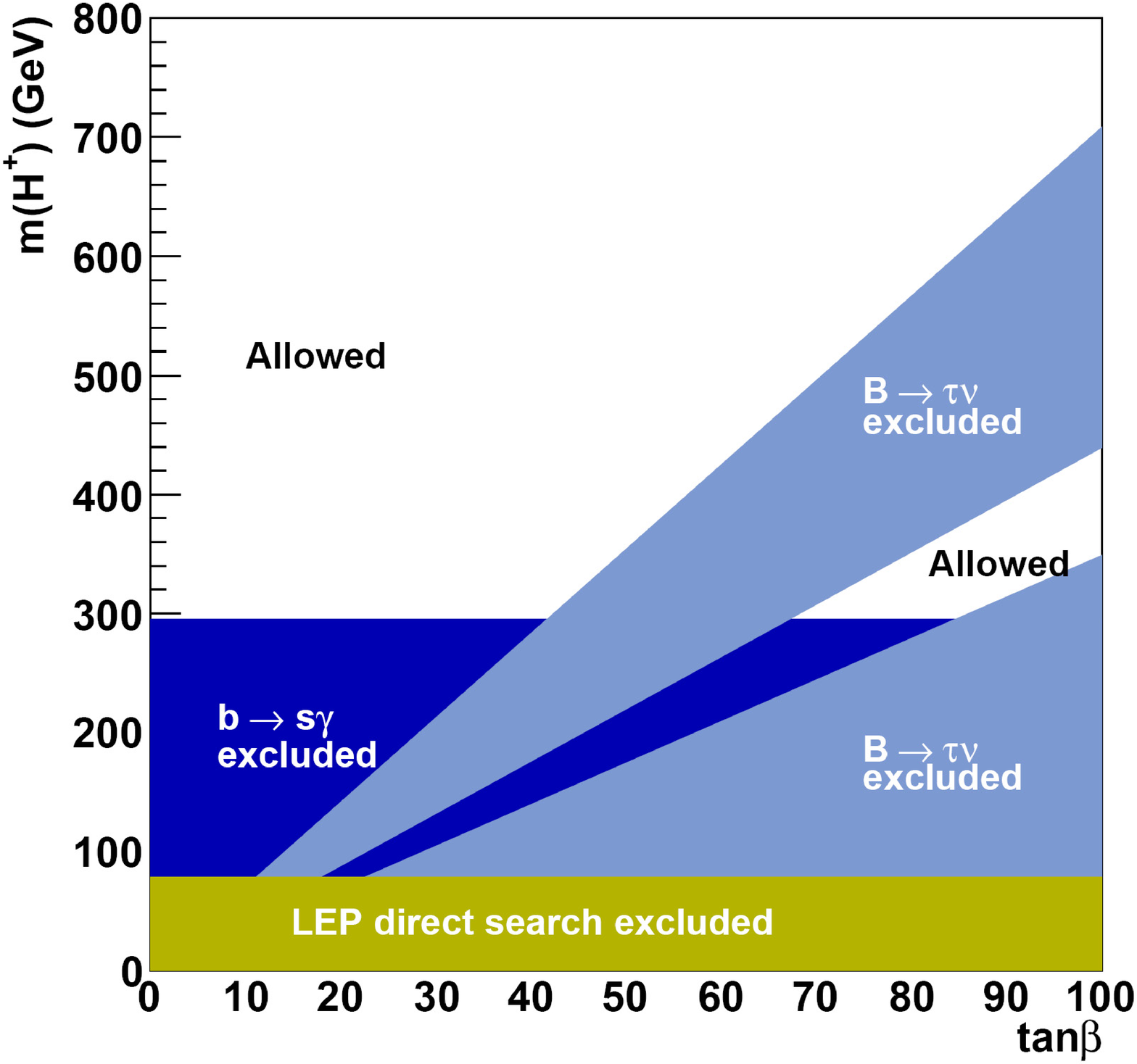}\label{low}
  \includegraphics[height=.35\textheight]{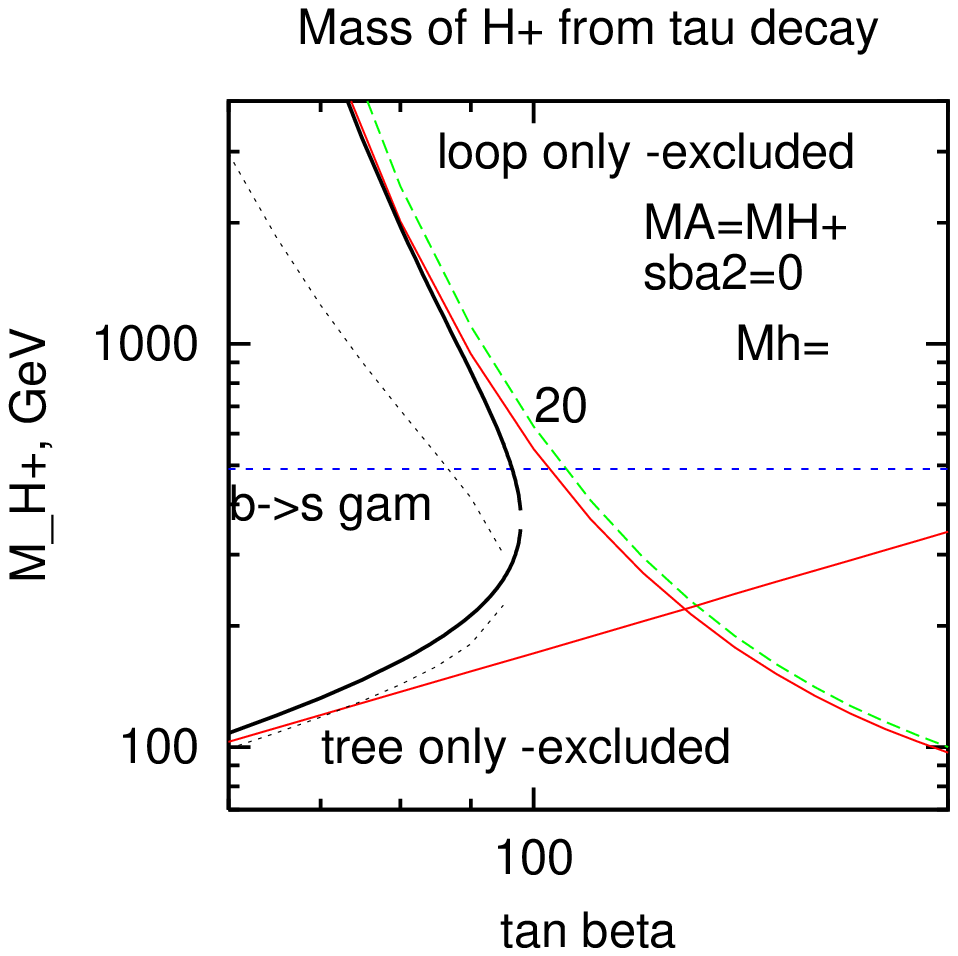}
   \caption{Left: Constraints from $B \to \tau \nu_\tau$ and $b \to s \gamma$ data on
    the charged Higgs boson  mass as a function of $\tan \beta$ in 2HDM
  (II)
  \cite{Du:2007wg}; Right: Limits from the leptonic $\tau$ decay for $M_h=20$ GeV and
  $\chi^h_{V}=0$ in 2HDM(II):
  tree-level exclusion of a region  below the straight line
   $M_{H^\pm} \geq$ 1.71 $\tan \beta$ {GeV} and
 one-loop exclusion of the region above the curve $ \Delta \sim \tan
\beta ^2\,\,[\ln \frac{M_{h}}{M_{H+}} + 1]$. The  excluded region
lies on the right on the curves: bold for $M_A=M_{H^\pm}$, dotted
for $M_A=100$ GeV. Exclusion from $\tau \to e\nu_\tau \bar{\nu_e}$
is represented by dashed line \cite{Krawczyk:2004na}.} \label{ch}
\end{figure}

Important constraints on mass of charged Higgs boson in 2HDM (II)
are coming from the $b \to s \gamma$ and $B \to \tau \nu$ decays.
The rate for the first process calculated at the NNLO accuracy in
the SM \cite{misiak}, after a comparison with the precise  data from
BaBar and Belle, leads to the constraint: $M_{H^\pm} > 295$ GeV at
95 \% CL for $\tan \beta
> 2$. This limit together with the constraints from the tree-level
analysis of $B \to \tau \nu$  \cite{Du:2007wg} is presented in
Fig.\ref{ch} (Left).

The 2HDM analysis has been performed at the one-loop level for  the
leptonic tau decays \cite{Krawczyk:2004na}. The constraints are
shown in Fig.\ref{ch} (Right). Not only lower, but also  in the
non-decoupling scenario upper limits can be derived here. In
contrast to the mentioned results from $b$ decays here the
(one-loop) constraints depend on masses of neutral Higgs bosons.

\section{Acknowledgment}
MK is grateful to Ilya Ginzburg and Rui Santos for important
discussions.

\begin{footnotesize}



%

\end{footnotesize}


\end{document}